\documentclass[twocolumn,showpacs,showkeys,preprintnumbers,prl,amsmath,amssymb]{revtex4}
\usepackage[dvips]{graphicx}
\usepackage{dcolumn}
\usepackage{bm}

\newcommand{\lk}{$L_k$ }
\newcommand{\lky}{$L_k$}
\newcommand{\lko}{$L_k(0)$ }

\newcommand{\lgg}{$L_g$ }
\newcommand{\lggy}{$L_g$}

\newcommand{\vtt}{$V(t)$ }
\newcommand{\vtty}{$V(t)$}
\newcommand{\itt}{$I(t)$ }
\newcommand{\itty}{$I(t)$}
\newcommand{\mmy}{$\mu$m}

\newcommand{\lamy}{$\lambda_L$}

\newcommand{\eqr}{Eq.~\ref}
\newcommand{\figr}{Fig.~\ref}

\newcommand{\aby}{$\sim$}

\newcommand{\tc}{$T_{c}$ }
\newcommand{\tcy}{$T_{c}$}

\newcommand{\be}{\begin{equation} }
\newcommand{\ene}{\end{equation}}

\newcommand{\jc}{$j_{c}$ }

\newcommand{\rhoo}{$\rho_{o}$ }
\newcommand{\rhooy}{$\rho_{o}$}

\newcommand{\ns}{$n_{s}$ }
\newcommand{\nsy}{$n_{s}$}

\begin{document}

\preprint{Published in Phys. Rev. Lett. {\bf 102}, 077001 (2009).}
\title{The ballistic acceleration of a supercurrent in a superconductor}

\author{Gabriel F. Saracila}
\author{Milind N. Kunchur} 
\email[Corresponding author email: ]{kunchur@sc.edu} 
\homepage{http://www.physics.sc.edu/kunchur}

\affiliation{Department of Physics and Astronomy, University of South
Carolina, Columbia, SC 29208}

\date{Submitted on 26 October 2008; accepted on 21 January 2009}

\begin{abstract}
One of the most primitive but elusive 
current-voltage (I-V) responses of
a superconductor is when its supercurrent grows steadily after a 
voltage is first applied. 
The present work employed a 
measurement system that could simultaneously track  and correlate $I(t)$
and $V(t)$ with sub-nanosecond timing accuracy, resulting in the first clear
time-domain measurement of this transient phase where the 
quantum system displays a Newtonian like response. The technique opens
doors for the controlled investigation of other time dependent transport
phenomena in condensed-matter systems. 
\end{abstract}


\maketitle

A particle under the action of a single applied force accelerates
ballistically in accordance with Newton's second law. 
In the presence of a frictional force, an applied force will 
ultimately maintain a constant velocity rather than produce acceleration.
Analogously, an externally applied voltage  $V$ maintains a constant
current $I$ in the case of a resistive conductor, whereas it 
can ballistically accelerate
the superfluid in a superconductor, leading to a supercurrent that grows
with time. 
This acceleration phase of the supercurrent lasts for a very brief 
period---until flux motion, phase slip centers, or other dissipative
phenomena set in---making it extremely difficult to observe
in the time domain in a correlated current-voltage measurement.
In the present work, an electrical measurement system was developed 
that could resolve and correlate the time evolutions of \itt and \vtt on a
subnanosecond time scale. For the sample pattern, extremely long meander
geometries (with length-to-width aspect ratios in the thousands) were
employed to prolong the acceleration time 
while maintaining $V$ at a 
manageable level. The combination of these two measures 
facilitated the successful observation of the acceleration phase. 

The acceleration of the supercurrent density $j_s$ is given 
by (from the London equations \cite{london,tinkhamtext})  
\be
\frac{dj}{dt} \simeq 
\frac{dj_s}{dt} = \frac{Ee^2n_s}{m^*} = \frac{E}{\mu_0 \lambda_L^2}, 
\label{LondonEq}
\ene
where $E$ is the local internal electric field, 
$e$ is the electronic charge, $m^*$ is the effective mass and 
\ns is the superfluid density (related
to the number of electrons per volume participating in the condensate);
the 
far right hand side of the equation relates \ns to the London magnetic-field 
penetration depth \lamy ; we can take
$j = j_s + j_n \simeq j_s$ because the normal current density $j_n$
is a negligible component of the total current density $j$.
This supercurrent acceleration phase lasts for 
the duration $\Delta t \approx j_c \mu_0 \lambda_L^2/E$, where 
\jc is the critical current density that marks the onset of resistance. 
The inductance-like proportionality between $dj/dt$ and $E$ in
\eqr{LondonEq},  arising
from the inertia of the superfluid, is referred to as the
kinetic inductance \lky . In terms of the 
geometrical length $l$ and cross sectional area $A$, it is given by 
\be
\label{lk_lambda}
L_k = \frac{\mu_0 \lambda^2 l}{A},
\ene
where $\lambda$ is a more general penetration depth, which includes effects
such as impurity scattering ($\lambda \geq \lambda_L$). 
Kinetic inductive effects are small except close to the transition temperature
\tcy , where their signatures have been seen in the high-frequency ac response
or as non-equilibrium inductive voltage spikes during abrupt current
steps \cite{oppenheim,geier,frank,anlage1,anlage2,lee,brorson,cho,jelila}. 
In the present work, timescales were chosen to be short enough to have a
sufficient magnitude of $V$ while long enough (compared to 
characteristic timescales such the gap-relaxation and
electron-phonon scattering times) to avoid non-equilibrium effects.
Variations in fields occured at length scales that were long compared
with both $\lambda$ and the coherence length $\xi$, so as to avoid non-local
effects. Thus the conditions were optimum for observing the simplest
limiting behavior of an accelerating condensate as predicted by the 
London equations, i.e., \eqr{LondonEq}. 

The samples used in this work were niobium films deposited on silicon 
substrates with DC magnetron sputtering. The films were patterned into
long narrow meanders by
electron-beam lithography using the lift-off technique. 
Sample A had a thickness of $t=70 \pm 8$ nm, a width of $w=12.1 \pm 0.6$
\mmy, and a length (between voltage probes) of $l=4.80 \pm 0.01$ cm. 
Sample B had the dimensions $t=85 \pm 8$ nm, $w=8.9 \pm 0.6$ \mmy, and
$l= 4.53 \pm 0.01$ cm. 
Their respective superconducting transition 
temperatures were $T_c$=6.74 K and $T_c$=7.23 K.

The measurements 
were carried out in a pulsed-tube closed-cycle refrigerator in zero
applied magnetic field.
The electrical measurements were conducted
with a pulsed signal source and detection electronics, in combination with 
a digital storage oscilloscope. Parts of the signal-source
and preamplifier circuitry in this setup 
were developed and built in-house. An active (buffered) ground 
arrangement was developed for improving the shielding between the fast
changing high-voltage signal in the current leads from the low-voltage
sample-signal sensing leads. The entire signal chain up until the digital
oscilloscope is analog. Using pulsed signals instead of 
continuous ac or dc excitations permits a wider 
range of currents without Joule heating of the sample and a 
flexible control over the waveform shape.
The system performs simultaneous independent differential 
four-probe measurements of $I(t)$ and $V(t)$ with a relative timing
accuracy of \aby 100 ps. 
The stray mutual inductive coupling between current and voltage leads 
has a (temperature independent) value of $\alt 1$ nH (the self
inductances of the leads themselves are not sensed because of the four-probe
configuration). The absolute accuracy of the inductance
values measured in this system is about $\pm$5\%. 
The voltage-measurement 
sensitivity is about 1 $\mu$V. The time interval between
digitized samples is $10^{-10}$ second (the single-shot digitizing
sampling rate is 10 GS/s). The speed and accuracy with which both \itt
and \vtt were tracked and correlated in a superconductor in the present
experiment are, to our knowledge, unprecedented. 

\begin{figure}[ht] 
\includegraphics[width=0.6\hsize]{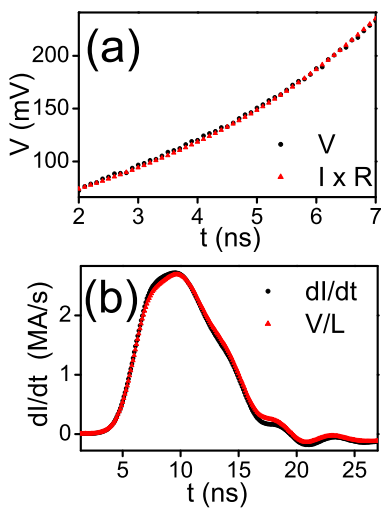}
\vspace{-1em}
\caption{{\em Measurement-apparatus temporal accuracy checks. (a) $V(t)$ and
$I(t)$ (multiplied by a constant $R$=62 $\Omega$) for a test resistor 
in place of a superconducting sample.  (b) $dI(t)/dt$ and $V(t)$ 
(divided by a constant $L$=15 nH) for a test inductor
in place of a superconducting sample. The voltages across the purely
resistive and inductive loads are seen to track their respective 
current and current-derivative functions 
 with sub-nanosecond accuracy.}}
\label{checks}
\end{figure}

Some tests and verifications of the measurement system are shown in 
\figr{checks}. Panel (a) shows the voltage and
current (scaled by a constant) for a purely resistive test sample and panel
(b) shows the current derivative and voltage (scaled by a constant) for
a purely inductive test sample. The time scales used in the actual
experiment were longer than these test conditions of \figr{checks} so that
the temporal tracking between the current and voltage sensing circuits 
was essentially perfect. Some additional information on the apparatus can be
found in our previous review articles \cite{mplb,pbreview}. 

\begin{figure}[ht] 
\includegraphics[width=0.98\hsize]{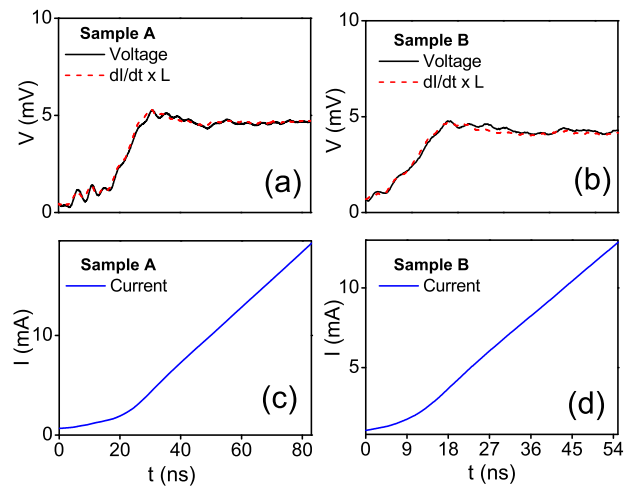}
\vspace{-1em}
\caption{{\em Time dependencies of voltage and  current 
in niobium meanders at T=3.84 K. Panels (a) and (b) represent 
\vtt and $dI/dt$ (scaled by a constant $L$) for samples A 
and B respectively. 
Panels (c) and (d)
show the corresponding  \itt functions, which rise 
steadily during the voltage plateaus of the panels above;
the time axes for panels (a) and (c) and for panels (b) and (d) are the
same.}} 
\label{IVcurves}
\end{figure}

Figs.~\ref{IVcurves}(a) and (b) 
show \vtt (solid lines) across two niobium-meander samples in the 
superconducting state at one temperature. 
Panels (c) and (d) show the corresponding \itt functions, which are 
seen to accelerate steadily during the plateaus in \vtty . 
The dashed lines in panels (a) and (b) show  $dI/dt$ scaled by a constant 
($L$=16.7 and 16.9 nH for samples A and B respectively) 
and are seen to track \vtt in instantaneous detail.
Thus the response is purely inductive, with an inductance that is 
independent of $I$ and $dI/dt$. 

\begin{figure}[ht] 
\includegraphics[width=0.98\hsize]{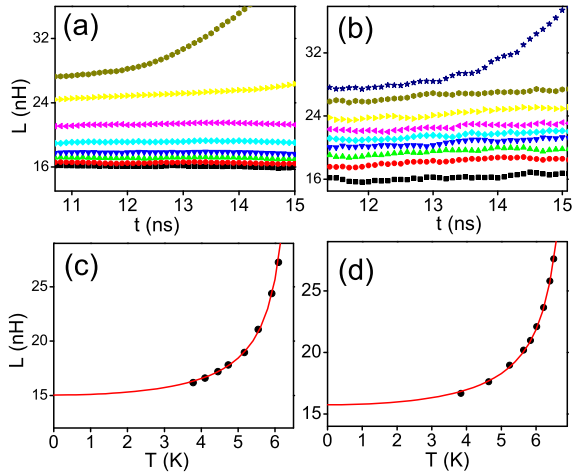}
\vspace{-1em}
\caption{{\em (a) Measured total inductance, $L(t) = V(t).dt/dI(t)$, versus
time for sample A. The 
curves correspond to the temperatures (from top to bottom): 
$T$ = 6.10, 5.91, 5.54, 5.17, 4.73, 4.10, and 3.78 K. 
(Each plot symbol on these curves corresponds to separately measured digital
voltage sample. The period between samples is 100 ps.)
(b) A similar $L(t)$ plot for sample B. For this panel, the temperatures
(from top to bottom) are: $T$ = 
 6.51, 6.4, 6.22, 6.02, 5.84, 5.24, 4.64, and 3.84 K.
Panels (c) and (d) show the temperature dependencies of the above 
measured $L$ values (taken on the plateaus around $t \sim 11$ ns) 
for samples A and B respectively. The symbols show the experimental data
and the solid line represents the least-squares fit to the 
two-parameter function 
 $L(T)=L_g + L_k(0)/[1-(T/T_c)^2]$ 
as discussed in the text.}}
\label{L-t-T}
\end{figure}

The ratio between the $dI(t)/dt$ and $V(t)$ curves in the top panels of
\figr{IVcurves} gives the time and current dependent inductance:
 $L(t) = V(t)/[dI(t)/dt]$. Figs.~\ref{L-t-T}(a) and (b) plot this 
 $L(t)$ versus time for various values of $T$, for each
of the samples. The plateau value of $L$ is seen to increase steadily
with temperature as is expected because of the declining superfluid density 
and consequent rising \lamy . Another interesting trend is
that the curves at highest temperatures show $L(t)$ functions that rise with
time (i.e., current). This happens because the current suppresses the
superfluid density through its pair-breaking action, 
a regime not seen before in any other kind of measurement. Note that
continuous-ac probes of \ns cannot endure high enough excitation levels to
explore this regime because of Joule heating; and tunneling
measurements reveal the spectral gap $\Omega_g$ rather than
\nsy . A systematic study of the suppression of \ns by $j$ 
will be the subject of a future investigation, since the
optimum sample geometry for studying this effect is almost opposite to
the sample geometry required for the present experiment.  

Figs.~\ref{L-t-T}(c) and (d) 
plot $L$ (as measured above) 
versus $T$ for each sample. 
This total inductance  $L$ =\lk +\lgg  has components
corresponding to the kinetic inductance \lk as well
as a geometrical inductance \lgg (from magnetic flux change). 
The temperature dependence arises chiefly from \lky ; the changes in \lgg with
temperature---arising from changes in the current-density profile across the
cross section---are relatively small (\aby 4\%). 
From \eqr {lk_lambda}
and the empirical temperature dependence of the penetration 
depth \cite{lee,brorson,cho} we have 
\be
\label{lk_T}
L_k (T) \approx \frac{L_k(0)}{[1-(T/T_c)^2]}, 
\ene 
 where $L_k(0)= \mu_0 \lambda^2(0) l/A$. 
The solid line curves in
\figr{L-t-T}(c) and (d) correspond to a least-squares fit to the 
function $L(T)=L_g + L_k(0)/[1-(T/T_c)^2]$.
The values of $L_k(0)$ and \lgg obtained from
this fitting (\tc is not a fitting parameter)
are listed in columns 2 and 4 of Table~\ref{summary-table}. The
coefficients of determination of the fits are 
$R^2$=0.9989 and $R^2$=0.9994 for samples A and B respectively. 
The standard errors of the fit combined with the error in the inductance 
measurement gives the error bars for \lky$(0)$ that are indicated in the table. 

\begin{table}
\begin{tabular}{||c|c|c|c|c||}
\hline \hline
& \multicolumn{2}{c}{$L_k(0)$  in nH}\vline&
\multicolumn{2}{c}{\lgg in nH} \vline \vline\\
\hline 
Sample&Expt.&Theor.&Expt.&Theor.\\
\hline
A&2.8$\pm$0.2&3.5$\pm$0.7&12.2$\pm$1&15.4$\pm$3\\
B&2.8$\pm$0.2&3.7$\pm0.7$&12.9$\pm$1&12.0$\pm$3\\
\hline \hline
\end{tabular}
\caption{{\em Experimentally observed values and theoretically estimated
values of the kinetic and geometrical inductances.}}
\label{summary-table}
\end{table}

The third and fifth columns of Table~\ref{summary-table} show,
for comparison, 
theoretical estimates of \lko and \lggy. 
For finding \lko we note that the effective penetration
depth ($\lambda$) becomes lengthened with respect to 
its clean-limit value ($\lambda_L$) 
in the presence of scattering by static disorder. 
This effect of impurity scattering can be incorporated
through the residual resistivity $\rho_o$ and expressed in terms of the 
order parameter $\Delta$ 
as \cite{TRL}
\be
\lambda(0) = \sqrt{\frac{\hbar\rho_o}{\pi \mu_o \Delta(0)}}. 
\ene
Taking the measured values of 
 \rhooy=0.347 $\mu \Omega$.m and 0.369  $\mu \Omega$.m,  
and obtaining $\Delta(0) \approx 2 k_B T_c$ from our measured values of
\tcy , we get $\lambda$=223 nm and 222 nm,
and \lky =3.5 nH and 3.7 nH  for samples A and B respectively. There is
an uncertainity in the values \rhoo because of the uncertainity in
sample dimensions and an uncertainity in  $\Delta(0)$ because of an uncertainity
 in the absolute value of \tc (this is roughly estimated to be around
200 mK). This gives rise to the error bars in the
theoretical \lk values that are stated in the table. The calculated 
values of  \lk are somewhat larger than the measured ones of but of
comparable magnitude. 

The theoretical geometrical inductances for the meanders, tabulated in the last
column of Table~\ref{summary-table}, were computed numerically 
by integrating the magnetic flux that links to the
path between the voltage probes. The error bars in the theoretical \lgg
arise from the uncertainities in the sample dimensions and the
approximations inherent in the calculation. 
It can be seen that the theoretically estimated \lgg values  
are also in agreement with their measured counterparts.

In summary, this work has explored the initial
acceleration phase during which a supercurrent builds up in response to an
applied voltage. 
The voltage and current curves of \figr{IVcurves} represent the first
clear and direct time-domain demonstration of this primitive regime, 
where the quantum system shows a Newtonian like response. It is also the
first time to observe the non-linear regime where the current
suppresses the superfluid density, thereby increasing the kinetic
inductance. The instrumentation developed for this experiment is unique and
represents the first measurement of its kind where both
\vtt and \itt are tracked in a superconductor with sub-nanosecond timing
accuracy. This technique can reveal more detailed information than just
an impulse-response measurement, and it 
can be used to explore time-dependent and
non-equilibrium phenomena in condensed-matter systems in a controlled
way (some examples of such regimes in superconductors would be those
related to phase slippage, glassy dynamics, and the nascent stage of a 
vortex right after its nucleation). 
The present work and its method should be distinguished from past experiments
in which an abrupt supercritical current pulse was 
applied \cite{frank,jelila} and only the subsequent \vtt
response was measured without monitoring \itty . In those experiments
the superconductor recoils in a highly  non-equilibrium manner to the
supercritical stimulus. In the present study, the superconducting system
is always maintained close to equilibrium by keeping the experimental
timescales well in excess of the gap-relaxation and electron-phonon 
scattering times, while keeping the timescales 
short enough to observe the inertia of the superfluid. 

The authors acknowledge useful discussions with 
J. M. Knight, B. I. Ivlev, R. A. Webb, R. J. Creswick, 
T. R. Lemberger, G. Simin, T. M. Crawford, and F. T. Avignone III. 
This research was supported by the U. S. Department of Energy 
through grant number DE-FG02-99ER45763.




\end{document}